\documentclass[%
 reprint,
 amsmath,amssymb,
pra,
]{revtex4-2}
\usepackage{booktabs}
\usepackage{multirow}
\usepackage{graphicx}
\usepackage{dcolumn}
\usepackage{bm}

\begin{document}

\preprint{APS/123-QED}

\title{Reference-frame-independent Quantum secure direct communication}
\author{Jia-Wei Ying,$^{1}$  Shi-Pu Gu,$^{1}$ Xing-Fu Wang,$^{3}$  Wei Zhong,$^4$  Ming-Ming Du,$^{1}$  Xi-Yun Li,$^{3}$  Shu-Ting Shen,$^{1}$  An-Lei Zhang,$^{3}$
	 Lan Zhou,$^{2,3}$}\email{zhoul@njupt.edu.cn}
\author{Yu-Bo Sheng$^{1,2}$}
 \email{shengyb@njupt.edu.cn}
\affiliation{%
 $^1$College of Electronic and Optical Engineering and College of Flexible Electronics (Future Technology), Nanjing
 University of Posts and Telecommunications, Nanjing, 210023, China\\
 $^2$School of Physics, Hangzhou Normal University, Hangzhou, Zhejiang 311121, China\\
 $^3$College of Science, Nanjing University of Posts and Telecommunications, Nanjing, 210023, China\\
 $^4$Institute of Quantum Information and Technology, Nanjing University of Posts and Telecommunications, Nanjing 210003, China
}%

\date{\today}

\begin{abstract}

Current quantum secure direct communication (QSDC) protocols guarantee communication security by estimating the error rates of photons in the X and Z bases. This take the reference frame calibration between communicating parties as a necessary prerequisite.
However, in mobile communications scenarios, achieving continuous and accurate reference frame calibration poses significant challenges. To address this issue, this paper proposes a reference-frame-independent (RFI) QSDC  protocol. This protocol only requires ensuring the calibration accuracy of one direction of the reference frame, while allowing a  misalignment angle $\beta$ in the other two directions.
To improve the protocol's robustness against reference frame fluctuations, we introduce a $\beta$-independent  parameter C into the security analysis framework and rederive the protocol's security bounds. Additionally,  we construct a system model and optimize the pulse intensity of the signal states, enabling the protocol to achieve optimal performance under each level of channel attenuation.
At an attenuation of 10 dB (corresponding to a communication distance of 25 km), the secrecy message capacities for $\beta= 0^{ \circ} $ and $45^{ \circ} $ are   $8.765 \times10^{-6}$ bit/pulse and $4.150 \times10^{-6}$ bit/pulse, respectively. Compared with the single-photon-based QSDC,  the communication distance of the protocol proposed in this paper is significantly extended. When  $\beta= 0^{ \circ} $ and $45^{ \circ} $, the maximum transmission distances of the RFI QSDC protocol are 27.875 km and 26.750 km, which is about 155.9\% and 149.7\% of that of the single-photon-based QSDC protocol.

\end{abstract}

\maketitle


\section{Introduction}

Quantum secure direct communication (QSDC) is a crucial branch of quantum communication. Owing to its ability to detect eavesdropping, it possesses unique potential advantages in communication scenarios. The concept of QSDC was first proposed by Long et al. \cite{QSDC1}, who later refined it to develop a two-step entanglement-based QSDC protocol \cite{QSDC2}. Over time, QSDC theory has witnessed substantial advancements, with the emergence of various protocols such as single-photon-based QSDC \cite{QSDC3}, measurement-device-independent (MDI) QSDC \cite{QSDC14n,QSDC24,MDI2}, passive-source QSDC \cite{pds,pdc,ps}, device-independent (DI) QSDC \cite{DI,QSDC18,QSDC21}, Receiver-device-independent (RDI) QSDC \cite{RDIQSDC}, one step QSDC \cite{QSDCos,QSDC16,QSDC16n}, and so on \cite{QSDChd,li,Masking,QSDCdense,QSDChy,QSDC22,QSDC23,QSDCpath,owQSDC}. 
Experimentally, the first single-photon-based QSDC demonstration was accomplished in 2016 \cite{QSDC9}, followed by subsequent breakthroughs including entanglement-based QSDC experiments \cite{QSDC10,QSDC11},  QSDC over 100 km fiber \cite{QSDC15},
QSDC network experiments \cite{QSDC12,QSDCnet}, continuous-variable QSDC experiments \cite{QSDC20n,cvexp}, and so on.

In a typical QSDC communication process, the message receiver first prepares and emits photons, which are then encoded with secret message by the message sender and sent back to the receiver. This process usually requires two rounds of security checking to ensure communication security and message accuracy.
Specifically, the first round is designed to detect eavesdropping and ensure the protocol’s security, while the second round verifies the recoverability of information to guarantee its correctness. 
For single-photon-based QSDC protocols, these two  rounds of security checking estimate the error rate by performing basis selection and measurement on single photons, and the error rate serves as a key criterion for judging whether the protocol operates normally. In detail, the second round of security checks is executed by the photon preparer (i.e., the message receiver), while the first round is conducted by the message sender.

Under ideal conditions, communication parties must maintain continuous reference frame alignment to ensure the accuracy of basis selection and measurement results. However, in some practical scenarios such as satellite-ground communication, the reference frame of one party  may undergo continuous changes with its movement, making it extremely difficult to achieve long-term stable alignment of reference frames.  Once a reference frame misalignment occurs, the error rate obtained from basis selection and measurement will be distorted, which in turn seriously impairs the security of the protocol.

The issue of reference frame misalignment also exists in quantum key distribution (QKD). In 2010, Laing et al. proposed reference-frame-independent QKD (RFI-QKD) to address this problem \cite{RFI10}. In their protocol, as long as one direction of the reference frame is aligned, the protocol can operate securely even if there is a certain misalignment angle $\beta$ in the other two directions. They introduced a  $\beta$-independent parameter $C$ to replace the error rate for estimating the protocol’s security. Since then, RFI-QKD has achieved significant progress both theoretically and experimentally\cite{RFIchip,RFIsou,RFIatmo,RFIpm,RFIMDI,RFIMDIfin,RFIml,RFI36,RFI44,RFI46,RFI6,RFIMDInn,RFITW}.

Inspired by previous work \cite{RFI10,RFI44,RFI6}, we introduce the  $\beta$-independent parameter $C$ into QSDC and propose a reference-frame-independent QSDC (RFI-QSDC) protocol. Our key contributions are as follows: 
(1) We provide a security proof for the RFI-QSDC protocol against collective attacks and photon-number-splitting (PNS) attacks, and rederive the formula for the secure message capacity of QSDC under reference frame misalignment.
(2)We construct a theoretical model and adopt the decoy state method to analyze the protocol’s performance in practical communication environments.
(3)We optimize the pulse intensity of the light source’s signal state, enabling the protocol to achieve optimal performance under each level of channel attenuation.

The structure of the paper is as follows. In Sec. II, we introduce  the RFI QSDC protocol. In Sec. III, we provide a security proof for the RFI-QSDC protocol and derive the formula for the secure message capacity of RFI QSDC. In Sec. IV, we establish a system model for the our protocol. In Sec. V, we give the decoy state method. In Sec. VI,  we conduct numerical simulation and parameter optimization. In Sec. VII, we give a  conclusion.

\section{The RFI QSDC protocol}

In this section, we will introduce the protocol steps of RFI QSDC. The steps of this protocol are similar to the  single-photon-based QSDC protocol \cite{QSDC3}, involving two participants: the message sender Alice and the message receiver Bob. 

Step 1: Initial state preparation.
First, Bob needs to prepare a series of photon pulses and randomly encode them into horizontal polarization state ($| H \rangle$), vertical polarization state ($| V \rangle$), diagonal  polarization state ($| + \rangle$), anti-diagonal polarization state ($| - \rangle$), right-handed circular polarization state ($| R \rangle$), or left-handed circular polarization state ($|L \rangle$) in a certain proportion. Among them, $| H \rangle$ and $| V \rangle$ states are classified  as the $Z_{B}$ basis, $|+ \rangle$ and $| - \rangle$ states are classified  as the  $X_{B}$ basis, and $| R \rangle$ and $| L \rangle$ states are classified  as the $Y_{B}$ basis.
 Then, Bob sends the sequence of photon pulses to Alice.
Assuming the initially prepared photon pulses are horizontally polarized photons ($| H \rangle$), the encoding operation can be expressed as:
\begin{align}
	M_{H}&=| H \rangle \langle H | +| V \rangle \langle V |,  \\ \nonumber 
	M_{V}&=| V \rangle \langle H | +| H \rangle \langle V |, \\ \nonumber 
	M_{+}&=| H \rangle \langle H | -| V \rangle \langle V |+| V \rangle \langle H | +| H \rangle \langle V |, \\ \nonumber 
	M_{-}&=| H \rangle \langle H | -| V \rangle \langle V |-| V \rangle \langle H | -| H \rangle \langle V |, \\ \nonumber 
	M_{R}&=| H \rangle \langle H | -| V \rangle \langle V |+i| V \rangle \langle H | -i| H \rangle \langle V |, \\ \nonumber 
	M_{L}&=| H \rangle \langle H | -| V \rangle \langle V |-i| V \rangle \langle H | +i| H \rangle \langle V |. \\ \nonumber 
\end{align}

Step 2: First round of security checking.
After receiving the photon sequence, Alice stores it. Bob publishes the positions of some $Z_{B}$ and $X_{B}$ basis  photons, as well as the positions of all  $Y_{B}$ basis photons. Alice selects the corresponding $Z_{A}$, $X_{A}$, and $Y_{A}$ basis for measurement to conduct  the first round of security check. In this process, we assume that Alice and Bob have only calibrated one direction ($Z$ basis) of their reference frames, while there is a misalignment  $\beta$ in the other two directions. The relationship between the bases used by Alice and Bob can be expressed as:
\begin{align}
	Z_{A}&=Z_{B}, \\ \nonumber
	X_{A}&=\cos\beta X_{B}+\sin\beta Y_{B}, \\ \nonumber
	Y_{A}&=\cos\beta Y_{B}-\sin\beta X_{B},
\end{align}
where $Z_{B}$, $X_{B}$, and $Y_{B}$  can be set as the three Pauli matrices $\sigma_{z}$, $\sigma_{x}$,
and $\sigma_{y}$, respectively.
Define $\xi_{B}\xi_{A}$ as follows: Bob prepares photons in the $\xi_{B}$ basis, and Alice chooses the $\xi_{A}$ basis for measurement, where  $\xi\in \{Z,X,Y\}$. Meanwhile, retain the measurement results of the five cases, i.e., $Z_{B}Z_{A}$, $X_{B}X_{A}$, $X_{B}Y_{A}$, $Y_{B}X_{A}$, and $Y_{B}Y_{A}$, which are used for parameter estimation to determine whether the security checking is passed. If passed, proceed to the next step; otherwise, terminate the protocol.

Step 3: Encoding.
Alice randomly selects some photons from the memory for random encoding, which is used for the second round of security checking. The remaining photons are encoded with information to transmit secret messages. Then send the encoded photons to Bob.
 Alice will perform two encoding operations $M_{0}$ and $M_{1}$. When performing operation $M_{0}$, the encoded information is recorded as 0; when performing operation $M_{1}$, the encoded information is recorded as 1. These two encoding operations can be expressed as:
\begin{align}
	M_{0}&=| H \rangle \langle H | +| V \rangle \langle V |,  \\ \nonumber 
	M_{1}&=| V \rangle \langle H | -| H \rangle \langle V |.
\end{align}

Step 4: Second round of security checking.
After receiving the photons, Bob performs measurements based on his initially prepared bases. Alice publishes the positions and encoding information of the randomly encoded photons. Bob compares the photons at the corresponding positions and conducts the second round of security checking to determine the correctness of the received information. If the security checking passes, proceed to the next step; otherwise, abandon  the received information.

Step 5: Message decoding.
Bob reconstructs the information sent by Alice based on the decoding information and the information of his initially prepared quantum states.

\section{ The theoretical secrecy message capacity}

According to Wyner's wiretap channel theory \cite{Wyner}, the secrecy message capacity of  QSDC can be defined as
\begin{align}
	C_{s}&=\max_{\{p_{0}\}} \{I(A:B)-I(A:E)\},
\end{align}
where $p_{0}$ is the probability that Alice encodes message $0$ with $M_{0}$ operation. According to information theory,
the maximum information capacity can be achieved when $p_{0}=0.5$.
$I(A:B)$ denotes the mutual information between Alice and Bob, which is the information entropy that Bob can obtain by decoding the photons encoded with information by Alice. Its value can be expressed as the product of the information entropy transmitted per pulse and  overall gain of the two-round transmission, and we have
\begin{eqnarray}
	I(A:B)=   Q^{BAB} [1-h( E^{BAB} )], 
\end{eqnarray}
where $Q^{BAB}$ is the overall gain, and  $ E^{BAB}$ is the quantum bit error rate. $h(x)$ is the binary Shannon entropy function, and  $h(x)=-x\log_{2}(x)-(1-x)\log_{2}(1-x)$.

$I(A:E)$  is the mutual information between Alice and Eve, representing the information entropy that Eve can obtain from Alice's encoded photons through attacks. Herein, we mainly consider collective attack and photon number splitting (PNS) attack.  $I(A:E)$ consists of two parts: the single-photon contribution and the multi-photon contribution, which can be written as
\begin{align}
	I(A:E)=	Q_{n=1 }^{BAE} I(A:E)_{n=1}+Q_{n\ge 2 }^{BAE} I(A:E)_{n\ge 2},
\end{align}
where $Q_{n=1 }^{BAE}$ and $Q_{n\ge 2 }^{BAE}$ are respectively the overall gains for Eve to intercept Alice's encoded photons. According to Ref. \cite{decoy3}, the overall gains of Eve can be bounded as
\begin{align}
	&Q_{n=1 }^{BAE}= P(I,1)(Y_{1}^{A}-Y_{0}^{A}),\\
	&Q_{n\ge 2}^{BAE}=Q^{BA}-Y_{0}^{A}-Q_{n=1 }^{BAE},
\end{align}
where $P(I,n)$ is the probability  that the source emits n-photon states with an intensity of $I$. $Y_{n}^{A}$ is the n-photon yield of Alice. 

$I(A:E)_{n=1}$ and  $I(A:E)_{n\ge 2}$ are respectively the information entropy that Eve can obtain from single-photon pulses and multi-photon pulses. For the sake of convenience, we consider the worst-case scenario and assume $I(A:E)_{n\ge 2}=1$.  Next, we will analyze the upper bound of the information entropy that Eve can obtain from single photons.
With the reference frame properly calibrated, the information entropy that Eve can steal from single-photon pulses has been given in Ref.  \cite{PQSDC1}, which is $h(e_{X}+e_{Z})$. $e_{X}$ ($e_{Z}$) is the phase (bit) error rate of single photons. Alice can estimate these values through the first round of security checking, thereby restricting Eve's eavesdropping behavior. However, when the reference frame cannot be calibrated at all times, $e_{X}$ and $e_{Z}$ will change due to the reference frame misalignment $\beta$, and Alice will then be unable to accurately estimate Eve's eavesdropping.

Inspired by Ref. \cite{RFI10}, we introduce $\beta$-independent invariants $C$ and $Q$, where 
\begin{align}
	Q&=\frac{1-\langle Z_{A}Z_{B} \rangle }{2}, \\
	C&=\langle X_{A}X_{B} \rangle^{2} +\langle X_{A}Y_{B} \rangle^{2} +
	\langle Y_{A}X_{B} \rangle^{2} +\langle Y_{A}Y_{B} \rangle^{2}.
\end{align}
$\langle \xi_{A}\xi_{B} \rangle$  represents the expected value when Bob prepares $\xi_{B}$ basis photons and Alice chooses the $\xi_{B}$ basis for measurement, $\langle \xi_{A}\xi_{B} \rangle=1-2e_{ \xi_{A}\xi_{B}}$ and $\xi\in \{Z,X,Y\}$. $e_{ \xi_{A}\xi_{B}}$ is the error rate when Bob prepares $\xi_{B}$ basis photons and Alice chooses the $\xi_{A}$ basis for measurement.

First of all, our protocol  is equivalent to Alice and Bob using an entangled state, where Bob measures one particle in the $X_{B}$, $Y_{B}$, or $Z_{B}$ basis and then sends the other particle to Alice.  Therefore, we assume that Alice and Bob share the entangled state $\rho _{AB}$. 
We can perform some operations to simplify the analysis, similar to the approach in Ref.  \cite{RFI10}.
Given that  C remains  invariant under the transformation $X_{A}\to -X_{A},Y_{A}\to -Y_{A},X_{B}\to -X_{B},Y_{B}\to -Y_{B} $,  and this transformation is equivalent to performing the $Z_{A}Z_{B} $ operation on $\rho _{AB}$, we can conduct an equivalent analysis of $ \hat{\rho} _{AB}=\frac{1}{2} (\rho _{AB}+Z_{A}Z_{B}\rho _{AB} Z_{A}Z_{B})$ and Eve can gain knowledge no less than that contained in  $\rho _{AB}$. In this way, we have
\begin{align}
	\hat{\rho} _{AB}=&\mu_{1}P_{\Phi _{+}}+\mu_{2}P_{\Phi _{-}}+
	(\frac{a}{2} |\Phi _{+}\rangle \langle\Phi _{-}| +H.c.)\\ \nonumber
	&+\mu_{3}P_{\Psi  _{+}}+\mu_{4}P_{\Psi  _{-}}+
	(\frac{b}{2} |\Psi  _{+}\rangle \langle\Psi_{-}|+H.c.),
\end{align}
where $P_{\Phi }=|\Phi\rangle \langle\Phi|$, and $|\Phi _{+}\rangle$, $|\Phi _{-}\rangle$, $|\Psi _{+}\rangle$, and $|\Psi _{-}\rangle$ represent the four standard orthogonal  Bell state. $H.c.$ represents the hermitian conjugation. Now, we have $C=2[(\mu_{1} -\mu_{2})^{2}  + (\mu_{3} -\mu_{4})^{2}  + Im(a)^{2}  +Im(b)^{2} ]  $. For convenience, let $\hat{\rho} _{AB}=\hat{\rho}(a,b)$. And $C$ is the same for $\hat{\rho}(-a^{\ast },-b^{\ast })$. Therefore, we can analyze $\rho^{'} =\frac{1}{2} (\hat{\rho}(a,b)+\hat{\rho}(-a^{\ast },-b^{\ast }))=\hat{\rho}(iA,iB)$, where $A=Im(a)$ and $B=Im(b)$. Finally, we can diagonalize $\hat{\rho}(iA,iB)$ as 
\begin{align}
	\rho^{'}_{AB} & = \sum_{k=1}^{4} \lambda _{k}|\psi_{k} \rangle \langle\psi_{k}|, 
\end{align}
where $\lambda _{1,2}=\frac{1}{2} (\mu_{1}+\mu_{2}\pm A^{'})$, $\lambda _{3,4}=\frac{1}{2} (\mu_{3}+\mu_{4}\pm B^{'})$,  $|\psi_{1,2}\rangle=\frac{1}{\sqrt{2} } (e^{i\chi }|HH\rangle \pm e^{-i\chi }|VV\rangle )$, and $|\psi_{3,4}\rangle=\frac{1}{\sqrt{2} } (e^{i\chi^{'} }|HV\rangle \pm e^{-i\chi^{'} }|VH\rangle )$. $A^{'}=\sqrt{(\mu_{1}-\mu_{2})^{2}+A^{2}  } $, $B^{'}=\sqrt{(\mu_{1}-\mu_{2})^{2}+B^{2}  } $, $\cos^{2}  \chi =\frac{1}{2} +\frac{\mu_{1}-\mu_{2}}{A^{'}} $, and $\cos^{2}  \chi^{'} =\frac{1}{2} +\frac{\mu_{3}-\mu_{4}}{B^{'}} $. Now, we have $C=2[(\lambda_{1} -\lambda_{2})^{2}  + (\lambda_{3} -\lambda_{4})^{2}]$.

Then, Eve can entangle his auxiliary system with Alice's and Bob's systems through a unitary operation and purify the system, resulting in
\begin{align}
	| \psi_{ABE}  \rangle =\sum_{k=1}^{4} \sqrt{\lambda _{k}} |\psi_{k} \rangle|E_{k} \rangle,
\end{align}
where $|E_{k} \rangle$ is a set of orthogonal states of Eve's auxiliary system.

Then Eve intercepts Alice's encoded photons and measures the  system AE. Herein, the  system AE can be obtained by taking the trace of the  system ABE, which is
\begin{align}
	\rho_{AE}^{'} =Tr_{B}(|\psi_{ABE}\rangle\langle\psi_{ABE}|)=\frac{1}{2}(P_{|\phi_{1}\rangle }+P_{|\phi _{2}\rangle}),
\end{align}
where 
\begin{align}
	|\phi _{1}\rangle=&e^{i\chi }| H \rangle (\sqrt{\lambda _{1}} |E_{1} \rangle +
	\sqrt{\lambda _{2}} |E_{2} \rangle)\\ \nonumber
	&+e^{-i\chi^{'} }| V \rangle (\sqrt{\lambda _{3}} |E_{3} \rangle-
	\sqrt{\lambda _{4}} |E_{4} \rangle), \\
	|\phi _{2}\rangle=&e^{i\chi^{'} }| H \rangle (\sqrt{\lambda _{3}} |E_{3} \rangle+
	\sqrt{\lambda _{4}} |E_{4} \rangle)\\ \nonumber
	&+e^{-i\chi }	| V \rangle (\sqrt{\lambda _{1}} |E_{1} \rangle-
	\sqrt{\lambda _{2}} |E_{2} \rangle).
\end{align}
After Alice completes the encoding operation, we have
\begin{align}
	|\phi _{1}\rangle=&e^{i\chi }| V \rangle (\sqrt{\lambda _{1}} |E_{1} \rangle +
	\sqrt{\lambda _{2}} |E_{2} \rangle)\\ \nonumber
	&-e^{-i\chi^{'} }| H \rangle (\sqrt{\lambda _{3}} |E_{3} \rangle-
	\sqrt{\lambda _{4}} |E_{4} \rangle), \\
	|\phi _{2}\rangle=&e^{i\chi^{'} }| V \rangle (\sqrt{\lambda _{3}} |E_{3} \rangle+
	\sqrt{\lambda _{4}} |E_{4} \rangle)\\ \nonumber
	&-e^{-i\chi }| H \rangle (\sqrt{\lambda _{1}} |E_{1} \rangle-
	\sqrt{\lambda _{2}} |E_{2} \rangle).
\end{align}
and
\begin{align}
	\rho_{AE,0} =\frac{1}{2}(P_{M_{0}|\phi_{1}\rangle }+P_{M_{0}|\phi _{2}\rangle}),\\ 
	\rho_{AE,1} =\frac{1}{2}(P_{M_{1}|\phi_{1}\rangle }+P_{M_{1}|\phi _{2}\rangle}),
\end{align}
where $\rho_{AE,0}$ and $\rho_{AE,1}$  denote the density matrices of the system after Alice encodes 0 and 1, respectively.

Finally, the system obtained by Eve can be expressed as
\begin{align}
	\rho ^{AE} =\frac{1}{2}(\rho_{AE,0} +\rho_{AE,1})
\end{align}

According to Holevo bound \cite{Holevo}, the upper bound of $I(A:E)_{n=1}$ can  be expressed as
\begin{align}
I(A:E)_{n=1}\le \max \{& S(\rho ^{AE})-\frac{1}{2}S(\rho _{AE,0})- \\ \nonumber
	&\frac{1}{2}S(\rho _{AE,1})\},
\end{align}
where $S(\rho)$ is von Neumann entropy, and $S(\rho)=-Tr(\rho \log_{2} \rho)=-\sum_{i}\lambda_{i}  \log_{2} \lambda_{i}$. $\lambda_{i}$ are the eigenvalues of $\rho$. 

In the bases $|\phi _{1}\rangle$ and $|\phi _{2}\rangle$, it can be easily obtained that $S(\rho_{AE,0})=S(\rho_{AE,1})=1$.
since the non-zero eigenvalues of $\rho ^{AE}$ are identical to those of its corresponding Gram matrix \cite{gram},  we can determine its von Neumann entropy by solving the Gram matrix.
The element in the $i$-th row and $j$-th column of the Gram matrix can be defined as:
\begin{align}
	G_{i,j}=\sqrt{p_{i}p_{j}}\langle \phi_{i}  | \phi_{j}  \rangle,
\end{align}
and we have
\begin{align}
	G=\frac{1}{4} \begin{pmatrix}
		1 & 0 &  0& x\\
		0& 1 & y & 0\\
		0 &-x  & 1 & 0\\
		-y &0  &0  &1
	\end{pmatrix},
\end{align}
where $x=-e^{-2i\chi }(\lambda_{1}-\lambda_{2})+e^{2i\chi^{'} }(\lambda_{3}+\lambda_{4})$, and $y=-e^{2i\chi }(\lambda_{1}-\lambda_{2})-e^{-2i\chi^{'} }(\lambda_{3}+\lambda_{4})$. The four eigenvalues of the Gram matrix are 
$\frac{1+\sqrt{-xy}}{4} $, $\frac{1+\sqrt{-xy}}{4}  $, $\frac{1-\sqrt{-xy}}{4}  $, and $\frac{1-\sqrt{-xy}}{4}  $.
Thus 
\begin{align}
	&S(\rho ^{AE} )  = h(\lambda_{Gram}) \\
	& = h(\frac{1+\sqrt{-xy}}{4} , \frac{1+\sqrt{-xy}}{4} , \frac{1-\sqrt{-xy}}{4} , \frac{1-\sqrt{-xy}}{4} ) 
	\\
	& = h(\frac{1+\sqrt{-xy}}{2})+1.
\end{align}

Since
\begin{align}
	xy=-(\lambda_{1}-\lambda_{2})^{2}-(\lambda_{3}-\lambda_{4})^{2}=-\frac{C}{2},
\end{align}
we have 
\begin{align}
I(A:E)_{n=1}\le& \max \{ S(\rho ^{AE})-\frac{1}{2}S(\rho _{AE,0})- \\ \nonumber
	&\frac{1}{2}S(\rho _{AE,1})\}\\ \nonumber
	= & h(\frac{1+\sqrt{\frac{C}{2}}}{2}).
\end{align}

Finally, the upper bound of the secrecy message capacity will be
\begin{align}
	C_{s}&=Q^{BAB} [1-h( E^{BAB} )]- \nonumber \\
	&\quad [Q_{s,n=1 }^{BAE}  h(\frac{1+\sqrt{\frac{C}{2}}}{2})+Q_{s,n\ge 2 }^{BAE}\cdot 1]. 
\end{align}

\section{System model}

To evaluate the performance of RFI QSDC in practical communication scenarios, we construct a system model to simulate it.
QSDC requires two rounds of transmission. We  define the first round of transmission as $BA$ and the second round as $BAB$. 
Consider a weak coherent light source with intensity $I$, which emits $n$-photon with a probability $P(I,n)$. This probability follows a Poisson distribution and $P(I,n)=e^{-I }\frac{I ^{n}}{n!}$.
In the protocol, Bob needs to emit six types of photons in the $X$, $Y$, and $Z$ bases. We may as well assume that the creation operators of the photons emitted by Bob is
\begin{align}
	a^{\dagger } (\theta , \phi ) & = \cos\frac{\theta }{2}a_{H}^{\dagger }+e^{i\phi}\sin\frac{\theta }{2}a_{V}^{\dagger }.
\end{align}
When $\theta=0, \pi$, they correspond to states $| H \rangle$ and $| V \rangle$.
When $\theta=\frac{\pi}{2}$ and $\phi=0, \pi, \frac{\pi}{2}, \frac{3\pi}{2}$, they correspond to states $| + \rangle$, $| - \rangle$, $| R \rangle$ and $| L \rangle$.

Assuming that the light source emits $n$ photons, which are reduced to  $k$ photons after channel loss, the probability of $k$ photons is
\begin{align}
	P_{k } &=  \sum_{n  =  k}^{\infty } P(I,n)S(n,k)\\ \nonumber
	&=\sum_{n  =  k}^{\infty } e^{-I }\frac{I ^{n}}{n!}C_{n}^{k} (1-\eta ^{chan})^{n-k}(\eta ^{chan})^{k}\\\nonumber
	&=\sum_{n  =  k}^{\infty } e^{-I }\frac{I ^{n}}{n!}\frac{n!}{(n-k)!k!}  (1-\eta ^{chan})^{n-k}(\eta ^{chan})^{k}\\\nonumber
	&\overset{t=n-k}{\rightarrow} \sum_{t  =  0}^{\infty }I^{k}e^{-I }\frac{I ^{t}}{t!k!}(1-\eta ^{chan})^{t}(\eta ^{chan})^{k}\\\nonumber
	&=\frac{(I\eta ^{chan})^{k}}{k!} e^{-I\eta ^{chan} }.
\end{align}
where $\eta ^{chan}$ is the transmission efficiency, and $\eta ^{chan}=t\eta_{opt}$. $t$ is the  channel transmission efficiency, and $t=10^{-\frac{\alpha L}{10}}$. $\alpha$ is the fiber attenuation coefficient and $L$ is the photon transmission distance. $\eta_{opt}$   is the intrinsic optical efficiency of the devices.

And we have
\begin{align}
	| k \rangle _{a^{\dagger } (\theta , \phi )}=&\frac{1}{\sqrt{k!} }
	(a^{\dagger } (\theta , \phi ))^{k} | vac \rangle   \label{x}\\ \nonumber
	=&\frac{1}{\sqrt{k!} }(\cos\frac{\theta}{2} a_{H}^{\dagger }+
	e^{i\phi  }\sin\frac{\theta}{2}a_{V}^{\dagger })^{k}| vac \rangle\\ \nonumber
	=&\frac{1}{\sqrt{k!} }(f(x) a_{x}^{\dagger }+
	f(y)a_{y}^{\dagger })^{k}| vac \rangle\\ \nonumber
	=& \frac{1}{\sqrt{k!} }\sum_{m=0}^{k} C_{k}^{m}(f(x)a_{x}^{\dagger })^{m}(f(y)a_{y}^{\dagger })^{k-m}| vac \rangle\\ \nonumber
	=&\frac{1}{\sqrt{k!} }\sum_{m=0}^{k} C_{k}^{m} f(x)^{m}
	f(y)^{k-m} \\ \nonumber
	&\sqrt{m!(k-m)!}  | m \rangle_{x}| k-m \rangle_{y},\\\nonumber
	(x,y)&\in  \{ (H,V),(+',-'),(R',L')  \}.
\end{align}

Since Alice needs to measure with her basis, we can expand the quantum state in terms of Alice's basis, that is
\begin{align}
	a^{\dagger } (\theta , \phi )=&\cos\frac{\theta}{2} a_{H}^{\dagger }+
	e^{i\phi  }\sin\frac{\theta}{2}a_{V}^{\dagger }    \label{a}
\\ \nonumber
	=&f(H)a_{H}^{\dagger }+f(V)a_{V}^{\dagger }\\ \nonumber
	=&\frac{1}{\sqrt{2} }(\cos\frac{\theta}{2}+e^{i(\phi -\beta )}\sin\frac{\theta}{2})a_{+'}^{\dagger }\\\nonumber
	&+\frac{1}{\sqrt{2} }(\cos\frac{\theta}{2}-e^{i(\phi -\beta )}\sin\frac{\theta}{2})a_{-'}^{\dagger }\\ \nonumber
	=&f(+')a_{+'}^{\dagger }+f(-')a_{-'}^{\dagger }\\ \nonumber
	=&\frac{1}{\sqrt{2} }(\cos\frac{\theta}{2}-ie^{i(\phi -\beta )}\sin\frac{\theta}{2})a_{R'}^{\dagger }\\\nonumber
	&+\frac{1}{\sqrt{2} }(\cos\frac{\theta}{2}+ie^{i(\phi -\beta )}\sin\frac{\theta}{2})a_{L'}^{\dagger }\\ 	\nonumber
	=&f(R')a_{R'}^{\dagger }+f(L')a_{L'}^{\dagger },
\end{align}
where $\beta$ is  the reference frame misalignment between Alice and Bob.
$a_{+'}^{\dagger }$ and $a_{-'}^{\dagger }$ ($a_{R'}^{\dagger }$ and $a_{L'}^{\dagger }$ ) are the creation operators of the eigenstates corresponding to $X_{A}$ ($Y_{A}$) basis. And 
$a_{+'}^{\dagger } = \frac{1}{\sqrt{2} } (a_{H}^{\dagger } +e^{i \beta } a_{V}^{\dagger })$,
$a_{-'}^{\dagger } = \frac{1}{\sqrt{2} } (a_{H}^{\dagger } -e^{i \beta } a_{V}^{\dagger })$,
$a_{R'}^{\dagger } = \frac{1}{\sqrt{2} } (a_{H}^{\dagger } +ie^{i \beta } a_{V}^{\dagger })$,
$a_{L'}^{\dagger } = \frac{1}{\sqrt{2} } (a_{H}^{\dagger } -ie^{i \beta } a_{V}^{\dagger })$.

According to Eq. (\ref{x}), the $k$-photon state arriving at Alice's end can be considered to generate state 
$| m \rangle_{x}| k-m \rangle_{y}$ with probability $P(k,m)$, and $P(k,m)=C_{k}^{m} |f(x)|  ^{2m}|f(y)| ^{2(k-m)}$.

Thus, the yield of $k$ photons arrive at detector and trigger response to state $x$ can be expressed as
\begin{align}
	Y_{k,x}&=\sum_{m=0}^{\infty } P(k,m)Y_{k,m} \\ \nonumber 
	&=\sum_{m=0}^{\infty }C_{k}^{m} |f(x)|  ^{2m}|f(y)| ^{2(k-m)} \\\nonumber 
	&\quad [1-(1-\eta_{D})^{m}(1-Pd)](1-\eta_{D})^{k-m}(1-Pd)\\\nonumber 
	&=(1-Pd)[ |f(x)|  ^{2}+|f(y)| ^{2}(1-\eta_{D})]\\\nonumber 
	&\quad -(1-\eta_{D}^{k})(1-Pd)^{2}\\\nonumber 
	&=(1-Pd)[ 1-|f(y)| ^{2}\eta_{D}]  -(1-\eta_{D}^{k})(1-Pd)^{2},
\end{align}
where $ \eta_{D}$  is the detection efficiency, and $Pd$ is the dark count probability.

Similarly,  the yield of $k$ photons arrive at detector and trigger response to state $Y$ can be expressed as
\begin{align}
	Y_{k,y}&=(1-Pd)[ 1-|f(x)| ^{2}\eta_{D}]  -(1-\eta_{D})^{k}(1-Pd)^{2}.
\end{align}

Then, the overall gain will be 
\begin{align}
	Q_{\xi_{A}\xi_{B}}&=Q_{x\xi_{B}}+Q_{y\xi_{B}}, \label{Q}
	\\ \nonumber
	Q_{x\xi_{B}}&=\sum_{k=0}^{\infty } P_{k}Y_{k,x} \\ \nonumber 
	&=(1-Pd)e^{-I\eta ^{chan}|f(y)|^{2}\eta_{D}}-(1-Pd)^{2}e^{-I\eta ^{chan} \eta_{D}},\\ \nonumber
	Q_{y\xi_{B}}&=\sum_{k=0}^{\infty } P_{k}Y_{k,y} \\ \nonumber 
	&=(1-Pd)e^{-I\eta ^{chan}|f(x)|^{2}\eta_{D}}-(1-Pd)^{2}e^{-I\eta ^{chan} \eta_{D}}.
\end{align}
Due to symmetry, we have
\begin{align}
	&Q_{Z_{A}Z_{B}}=Q_{HZ_{B}}+Q_{VZ_{B}}=Q_{HH}+Q_{VH}, \label{qxx}\\ \nonumber
	&	Q_{X_{A}X_{B}}=Q_{+^{'}X_{B}}+Q_{-^{'}X_{B}}=Q_{+^{'}+}+Q_{-^{'}+},\\\nonumber
	&Q_{X_{A}Y_{B}}=Q_{+^{'}Y_{B}}+Q_{-^{'}Y_{B}}=Q_{+^{'}R}+Q_{-^{'}R},\\\nonumber
&	Q_{Y_{A}X_{B}}=Q_{R^{'}X_{B}}+Q_{L^{'}X_{B}}=Q_{R^{'}+}+Q_{L^{'}+},\\\nonumber
&Q_{Y_{A}Y_{B}}=Q_{R^{'}Y_{B}}+Q_{L^{'}Y_{B}}=Q_{R^{'}R}+Q_{L^{'}R},
\end{align}
where $Q_{\xi_{A}\xi_{B}}$ denotes the scenario where  Bob emits a state in $\xi_{B}$ basis and Alice measures with the $\xi_{A}$ basis. $Q_{x\xi_{B}}$  refers to the case  where Bob emits a state in the $\xi_{B}$  basis, and Alice obtains an $x$ response in the $\xi_{A}$ basis. $Q_{xy}$ represents the situation where Bob emits the state $y$ in the $\xi_{B}$  basis and Alice obtains a $x$ response in the $\xi_{A}$ basis.

According to Eq. (\ref{a}), we have
\begin{align}
	&|f(H)| ^{2}=\frac{1+\cos\theta }{2} ,|f(V)| ^{2}=\frac{1-\cos\theta }{2} ,\\ \nonumber
	&|f(D')| ^{2}=\frac{1+\sin\theta \cos(\phi-\beta)}{2} ,\\ \nonumber
	&|f(A')| ^{2}=\frac{1-\sin\theta \cos(\phi-\beta)}{2}, \\ \nonumber
	&|f(R')| ^{2}=\frac{1+\sin\theta \sin(\phi-\beta)}{2} ,\\ \nonumber
	&|f(L')| ^{2}=\frac{1-\sin\theta \sin(\phi-\beta)}{2},
\end{align}
and
\begin{align}
	Q_{HH}&=(1-Pd)[1-(1-Pd)e^{-I\eta^{chan}\eta_{D}}], \label{qxy}\\ \nonumber 
	Q_{VH}&=(1-Pd)Pd e^{-I\eta^{chan}\eta_{D}}, \\ \nonumber 
	Q_{+^{'}+}&= Q_{R^{'}R}      \\ \nonumber 
	&=(1-Pd)[e^{-I\eta^{chan}\eta_{D}\frac{1+\cos\beta }{2} }-(1-Pd)e^{-I\eta^{chan}\eta_{D}}], \\ \nonumber 
	Q_{-^{'}+}&= Q_{L^{'}R}      \\ \nonumber 
	&=(1-Pd)[e^{-I\eta^{chan}\eta_{D}\frac{1-\cos\beta }{2} }-(1-Pd)e^{-I\eta^{chan}\eta_{D}}], \\ \nonumber 
	Q_{+^{'}R}&= Q_{R^{'}-}      \\ \nonumber 
	&=(1-Pd)[e^{-I\eta^{chan}\eta_{D}\frac{1+\sin\beta }{2} }-(1-Pd)e^{-I\eta^{chan}\eta_{D}}], \\ \nonumber 
	Q_{+^{'}L}&= Q_{R^{'}+}      \\ \nonumber 
	&=(1-Pd)[e^{-I\eta^{chan}\eta_{D}\frac{1-\sin\beta }{2} }-(1-Pd)e^{-I\eta^{chan}\eta_{D}}].
\end{align}

According to Eq. (\ref{Q}), the error  rate of state $| x \rangle$ can be expressed as
\begin{align}
	E_{\xi_{A}\xi_{B}}=E_{x}=\frac{e_{d}Q_{x\xi_{B}}+(1-e_{d})Q_{y\xi_{B}}}{Q_{x\xi_{B}}+Q_{y\xi_{B}}} ,
\end{align}
where $e_{d}$ is the optical intrinsic error rate. According to Eq. (\ref{qxy}), we can  obtain the expression for the error rate.

\section{Decoy state method}
In the protocol, we need to obtain the error rate of single-photon pulses ($e_{\xi_{A}\xi_{B}}$) to calculate $Q$ and  $C$ . However, these parameters cannot be directly measured in practical scenarios, so the decoy state method needs to be used for estimation. By solving the following linear programming problem, we can obtain the estimated value of them.
\begin{align}
	\min  &\quad  Y_{1,\xi_{A}\xi_{B}}     \\ \nonumber
	s.t.&\quad Q_{\xi_{A}\xi_{B}}^{I}   \ge \sum_{n=0}^{n_{cut}}  P(I,n)   Y_{n,\xi_{A}\xi_{B}}    ,\\  \nonumber
	& \quad Q_{\xi_{A}\xi_{B}}^{I}   \le \sum_{n=0}^{n_{cut}}  P(I,n)    Y_{n,\xi_{A}\xi_{B}}
	+1-\sum_{n=0}^{n_{cut}}  P(I,n)  ,\\  \nonumber
	&0\le Y_{n,\xi_{A}\xi_{B}}    \le 1 ,
\end{align}
\begin{align}
	\max & \quad  e_{1,\xi_{A}\xi_{B}}  Y_{1,\xi_{A}\xi_{B}}\\  \nonumber
	s.t.&\quad  Q_{\xi_{A}\xi_{B}}^{I} E_{\xi_{A}\xi_{B}}^{I}\ge \sum_{n=0}^{n_{cut}}P(I,n) e_{n,\xi_{A}\xi_{B}}Y_{n,\xi_{A}\xi_{B}}   ,\\  \nonumber
	& \quad Q_{\xi_{A}\xi_{B}}^{I} E_{\xi_{A}\xi_{B}}^{I}  \le \sum_{n=0}^{n_{cut}}   P(I,n)    e_{n,\xi_{A}\xi_{B}} Y_{n,\xi_{A}\xi_{B}} \\  \nonumber
	&\quad\quad\quad\quad\quad\quad\quad+1-\sum_{n=0}^{n_{cut}}   P(I,n) ,\\  \nonumber
	& 0\le   e_{n,\xi_{A}\xi_{B}} Y_{n,\xi_{A}\xi_{B}}   \le 1 .
\end{align}
where $I$ denotes the intensity of decoy states and signal state.

\section{Numerical simulation}
In this section, we conduct numerical simulations of the system to evaluate the performance of our protocol. The parameters used in the simulations are derived from Ref. \cite{decoy3} and are detailed in Table \ref{para}.
\begin{table}[h]
	\centering
	\vspace{-0.1cm}
	\setlength{\abovecaptionskip}{0.3cm}
	\setlength{\belowcaptionskip}{0.1cm}
	\setlength\tabcolsep{5pt} 
	\renewcommand\arraystretch{1.5}  
	\caption{Parameters used in the numerical simulation \cite{decoy3}.}
	\begin{tabular}{cccccc}  
		\hline
		\hline
		$\eta_{opt}^{BA}$ & $\eta_{opt}^{BAB}$ & Pd &$e_{d}^{A}$   & $e_{d}^{B}$   & $\eta_{D}$\\
		0.21    & 0.088    & $8 \times 10^{-8}$  &0.0131     & 0.0026    & 0.7  \\  \hline \hline
	\end{tabular}\label{para}
\end{table}

\begin{figure}[!htbp]\label{dis}
	\begin{center}
		\includegraphics[width=8cm,angle=0]{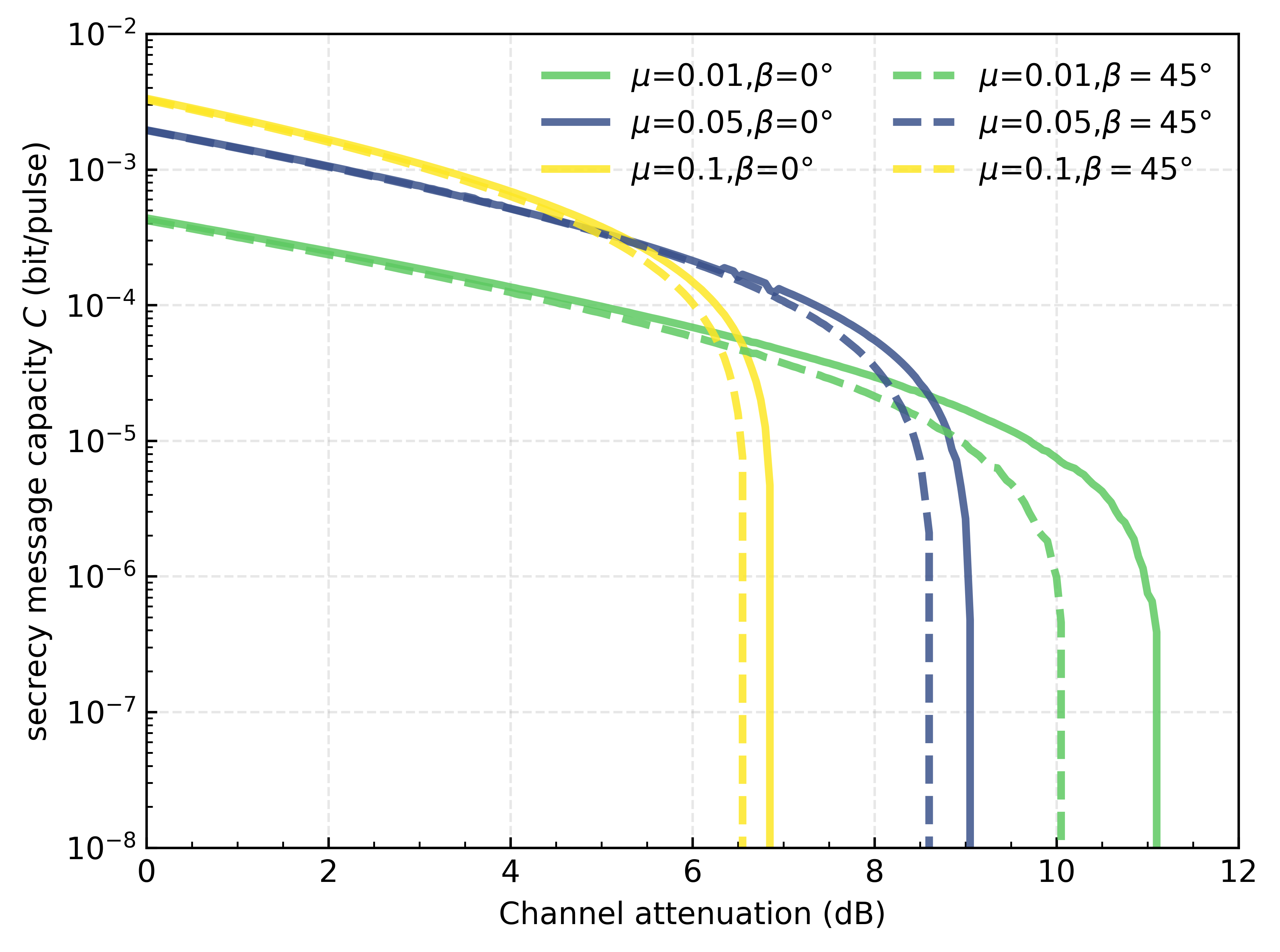}
		\caption{The relationship between channel attenuation and secrecy message capacity of the protocol under different pulse intensities of the light source's signal state. The solid (dashed) lines in yellow, blue, and green correspond to the scenarios where the signal state intensities are 0.1, 0.05, and 0.01 with $\beta$=$0^{ \circ} $ ($45^{ \circ} $), respectively.
		}\label{simulation1}
	\end{center}
\end{figure}

In Fig. \ref{simulation1}, we plot the curves illustrating the relationship between channel attenuation and secrecy message capacity of the protocol under different pulse intensities of the light source's signal state. The intensities of the signal state and decoy states are set as follows: $I_{s}=\mu$, $I_{d1}=0.05\mu$, and $I_{d2}=0.01\mu$, where the subscripts 's', 'd1', and 'd2' denote the signal state, decoy state 1, and decoy state 2, respectively.
In the figure, the solid lines in yellow, blue, and green correspond to the scenarios where the signal state intensities are 0.1, 0.05, and 0.01 (with $\beta$=$0^{ \circ} $), respectively. Additionally, we also plot the curve for a reference frame misalignment of $45^{ \circ} $ (represented by the dashed line in the figure).
Under ideal conditions, the performance of the protocol should be associated with the $\beta$-independent parameter C. However, since the decoy state method cannot accurately estimate the gain and error rate, the estimated value of C exhibits a $\beta$-dependent periodicity. When $\beta=45^{ \circ}$, C reaches its minimum value, and at this point, the information entropy that Eve can steal reaches its maximum. In other words, the actual performance of the protocol will lie between the solid lines and the dashed line depending on the actual value of $\beta$.

\begin{figure}[!htbp]
	\begin{center}
		\includegraphics[width=8cm,angle=0]{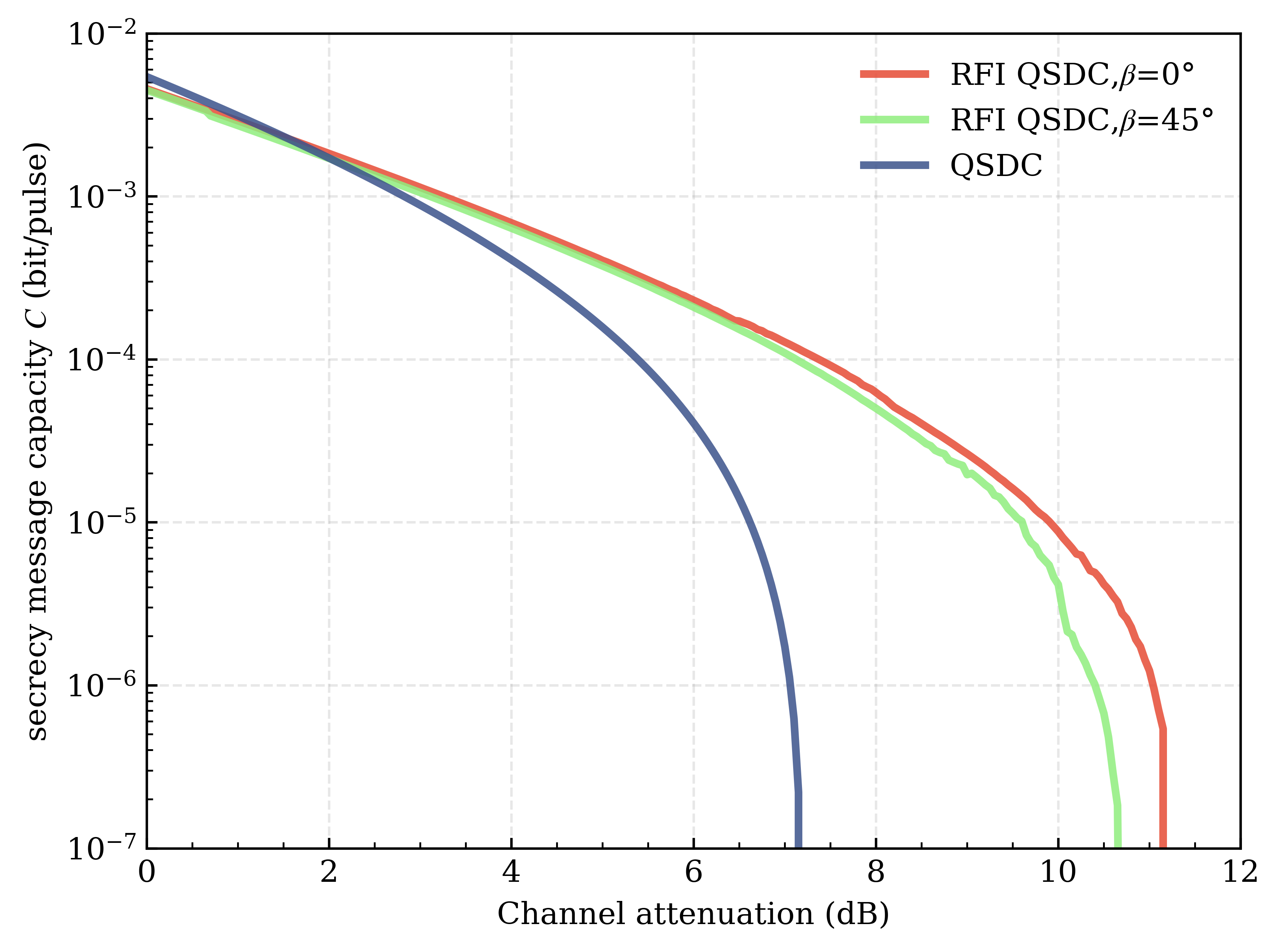}
		\caption{Comparison of secrecy message capacity at the optimized intensity between RFI QSDC and single-photon-based QSDC Protocols. The red and green curves correspond to the RFI QSDC protocol with reference frame misalignment of $0^{ \circ} $ and $45^{ \circ} $, respectively, while the blue curve represents the single-photon-based QSDC protocol.
		}\label{com}
	\end{center}
\end{figure}

From Fig. \ref{simulation1}, we can observe that under different channel attenuations, the secrecy message capacities corresponding to the three intensities vary (some higher, some lower). To obtain the maximum secrecy message capacity of the protocol for each channel attenuation, we optimized the light source intensity of the protocol.
In Fig. \ref{com}, we plot the curves showing the relationship between the secrecy message capacity (at the optimized intensity) and channel attenuation for the protocol when the reference frame misalignment is $0^{ \circ} $ and $45^{ \circ} $. We also compare this with the single-photon-based QSDC protocol under its optimal parameters. Among the curves, the red and green ones correspond to the RFI QSDC protocol with reference frame misalignment of $0^{ \circ} $ and $45^{ \circ} $, respectively, while the blue curve represents the single-photon-based QSDC protocol under optimal parameters.
When the channel attenuation is low, the impact of different reference frame misalignment values on the secrecy message capacity is not significant; however, this impact becomes increasingly prominent as the channel attenuation increases. For the protocol with $\beta= 0^{ \circ} $ and $45^{ \circ} $, the maximum channel attenuations are 11.15 dB and 10.7 dB, respectively (corresponding to maximum communication distances of 27.875 km and 26.750 km, respectively). At an attenuation of 10 dB (corresponding to a communication distance of 25 km), the secrecy message capacities for $\beta= 0^{ \circ} $ and $45^{ \circ} $ are   $8.765 \times10^{-6}$ bit/pulse and $4.150 \times10^{-6}$ bit/pulse, respectively.

Compared with the single-photon-based QSDC, our protocol performs slightly lower at low attenuation. This is mainly because we only consider the message transmitted by single-photon pulses while neglecting the contribution of two-photon components. In the original QSDC protocol, two-photon pulses can also make a positive contribution to the secrecy message capacity.
As the transmission distance increases, the optimal pulse intensity gradually decreases, and the proportion of two-photon components reduces—at this point, the performance of the RFI QSDC quickly surpasses that of the single-photon-based QSDC. At a channel attenuation of 6 dB (corresponding to a transmission distance of 15 km), the secrecy message capacities are as follows: for the RFI QSDC with  $\beta= 0^{ \circ} $ and $45^{ \circ} $, the values are $2.304 \times10^{-4}$ bit/pulse and $2.089 \times10^{-4}$ respectively; for the single-photon-based QSDC, the value is $4.059 \times10^{-5}$ bit/pulse.
Additionally, the maximum transmission distance of the RFI QSDC protocol is also far greater than that of the single-photon-based QSDC protocol. When  $\beta= 0^{ \circ} $ and $45^{ \circ} $, the maximum transmission distances of the RFI QSDC protocol are 10 km and 8.875 km longer than that of the single-photon-based QSDC protocol, with improvement rates of 55.9\% and 49.7\%, respectively.

\section{Conclusion}
In this paper, we propose a RFI QSDC  protocol. To address the issue of reference frame misalignment, we introduce a $\beta$-independent  parameter C into the security analysis framework of QSDC, and rederive the security bounds of the proposed protocol under collective attacks and PNS attacks. We improve the secrecy message capacity formula of QSDC: the original calculation method, which relied on both the X-basis and Z-basis error rates, is updated to a new form that only depends on parameter C.
In scenarios where the reference frame undergoes dynamic changes, the protocol  only requires ensuring the calibration accuracy of one direction of the reference frame, and secrecy message can still be achieved even if there are reference frame misalignment deviations in the other two directions. To verify the protocol performance, we construct a corresponding system model. We optimize the pulse intensity of the signal states, enabling the protocol to achieve optimal performance under each level of channel attenuation.
At an attenuation of 10 dB (corresponding to a communication distance of 25 km), the secrecy message capacities for $\beta= 0^{ \circ} $ and $45^{ \circ} $ are   $8.765 \times10^{-6}$ bit/pulse and $4.150 \times10^{-6}$ bit/pulse, respectively. Compared with the single-photon-based QSDC,  the communication distance of the protocol proposed in this paper is significantly extended. When  $\beta= 0^{ \circ} $ and $45^{ \circ} $, the maximum transmission distances of the RFI QSDC protocol are 27.875 km and 26.750 km, which is about 155.9\% and 149.7\% of that of the single-photon-based QSDC protocol.
This protocol demonstrates broad application prospects in future mobile communication scenarios, providing a potential solution for the implementation and promotion of QSDC in this field.

\appendix

\section*{Acknowledgement}
 This work is supported by the National Natural Science Foundation of China under Grants No. 12574393, No.92365110, and No. 12175106, and the Postgraduate Research $\&$ Practice Innovation Program of Jiangsu Province under Grant No. KYCX25-1129.

\end{document}